\documentclass[%
aps,%
prb,%
twocolumn,%
showpacs]%
{revtex4}
\usepackage{graphicx}% Include figure files
\usepackage{bm}% bold math
\usepackage{color}%
\usepackage{amsmath,amssymb}

\newcommand{\molecule}[1]{\ensuremath{\mathrm{#1}}}

\newcommand{\tmfloatsmallb}[2]{
\begin{figure}[b]
#1
\caption{#2}
\end{figure}}

\newcommand{\tmfloatsmall}[2]{
\begin{figure}
#1
\caption{#2}
\end{figure}}

\newcommand{\tmtablesmall}[2]{
\begin{table}
#1
\caption{#2}
\end{table}}

\begin{document}

\title{Benchmark all-electron \textit{ab initio} quantum Monte Carlo
calculations for small molecules}

\author{Norbert Nemec}
\author{Michael D. Towler}
\author{R. J. Needs}
\affiliation{TCM Group, Cavendish Laboratory,
 University of Cambridge,
 J.J. Thomson Ave., CB3 0HE, United Kingdom}

\date{\today}

\begin{abstract}
  We study the efficiency, precision and accuracy of all-electron variational
  and diffusion quantum Monte Carlo calculations using Slater basis sets.
  Starting from wave functions generated by Hartree-Fock and density
  functional theory, we describe an algorithm to enforce the electron-nucleus
  cusp condition by linear projection. For the 55 molecules in the G2 set, the
  diffusion quantum Monte Carlo calculations recovers an average of 95\% of
  the correlation energy and reproduces bond energies to a mean absolute
  deviation of 3.2~kcal/mol. Comparing the individual total energies with
  essentially exact values, we investigate the error cancellation in
  atomization and chemical reaction path energies, giving additional insight
  into the sizes of nodal surface errors.
\end{abstract}


\maketitle

\section{Introduction}

Ab initio variational Monte Carlo (VMC) and diffusion Monte Carlo (DMC)
methods have been used successfully for systems containing hundreds and
sometimes thousands of electrons.\cite{needs-cvadqmcc2010} Such
calculations typically retrieve 95\% or more of the correlation energy within
the fixed-node approximation based on a single determinant. Testing the
accuracy of VMC and DMC results against experiment and other theoretical
methods plays an important role in the development of computer codes such as
the CASINO package.\cite{needs-cvadqmcc2010}

As a method that directly competes for accuracy and efficiency with
deterministic quantum chemical methods, benchmark results are of great
interest. Total electronic energies of atoms lend themselves to direct
comparison as highly accurate reference values are available. DMC calculations
can recover 99\% or more of the correlation energy for first row atoms using
multi-determinant and backflow wave functions.\cite{brown-eotfrafqmc2007}
Another well-known set of benchmark data are the atomization energies of the
G2 set of molecules.\cite{curtiss-gtfmeofasc1991} These energies have been
reproduced in DMC to high accuracy using
pseudopotentials.\cite{grossman-bqmcc2002} Calculations of a selection of
small molecules consisting of first-row atoms have proven all-electron DMC to
be nearly as accurate as CCSD(T)/cc-pVTZ.\cite{manten-otaotfdqmcm2001}

In every case, the dominant deviations were attributed to the fixed node
approximation made in the DMC method. DMC satisfies the variational principle
and therefore the energies are too high when an approximate nodal surface is
used. Comparing total energies for benchmarking purposes, DMC typically
performs extremely well. For atomization and chemical reaction energies,
however, error cancellation in the energy differences becomes essential.
Unfortunately, the quality of the nodal surface turns out to depend
intricately on the chemical structure of each molecule or atom and the error
cancelation in DMC is less efficient and predictable than in competing
methods.

\tmfloatsmallb{\resizebox{80mm}{!}{\includegraphics{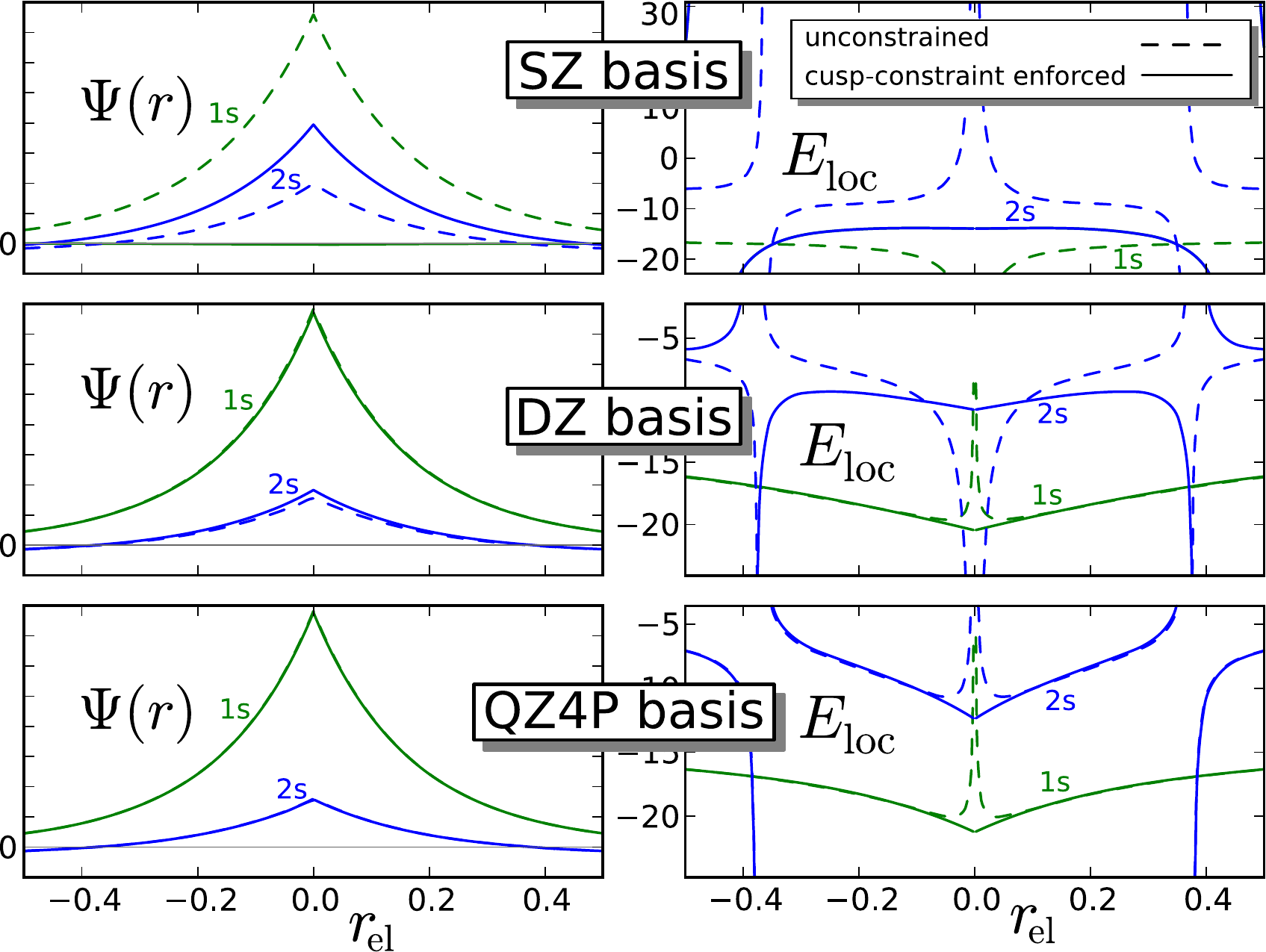}}}{\label{fig:cusp-correction}Effect
of the nuclear cusp constraint on orbitals $\Psi (r)$ (left) and orbital local
energies $E_{\operatorname{loc}} = \Psi^{- 1} \mathcal{H} \Psi$ (right) for the carbon
atom. Three different general-purpose basis sets from the ADF package were
used (top to bottom with increasing size and precision). For atoms, only s
orbitals require cusp correction and are here shown before and after the
constraint for comparison. Within the single-$\zeta$ basis SZ the wave
function is severely distorted when the coefficient of the single 1s basis
function is adjusted; for the double-$\zeta$ basis DZ, the local energy is
still strongly distorted; for the quadruple-$\zeta$ basis QZ4P, the divergence
in the local energy at the nucleus is cleanly removed, otherwise preserving
the orbitals and the local energy.}

In this paper we present benchmark results for the aforementioned 55 molecules
of the G2 set from all-electron DMC calculations using wave functions based on
Slater-type orbitals. These results are directly comparable to
pseudopotential-based calculations. Beyond this, however, the availability of
highly accurate reference data for the total energies of the same set of
molecules\cite{oneill-bcefsm2005} allows a deeper analysis of the relative
nodal surface errors in various chemical species and permits an intuitive
visualization of error cancelation in chemical reaction energies.

\tmfloatsmall{\resizebox{80mm}{!}{\includegraphics{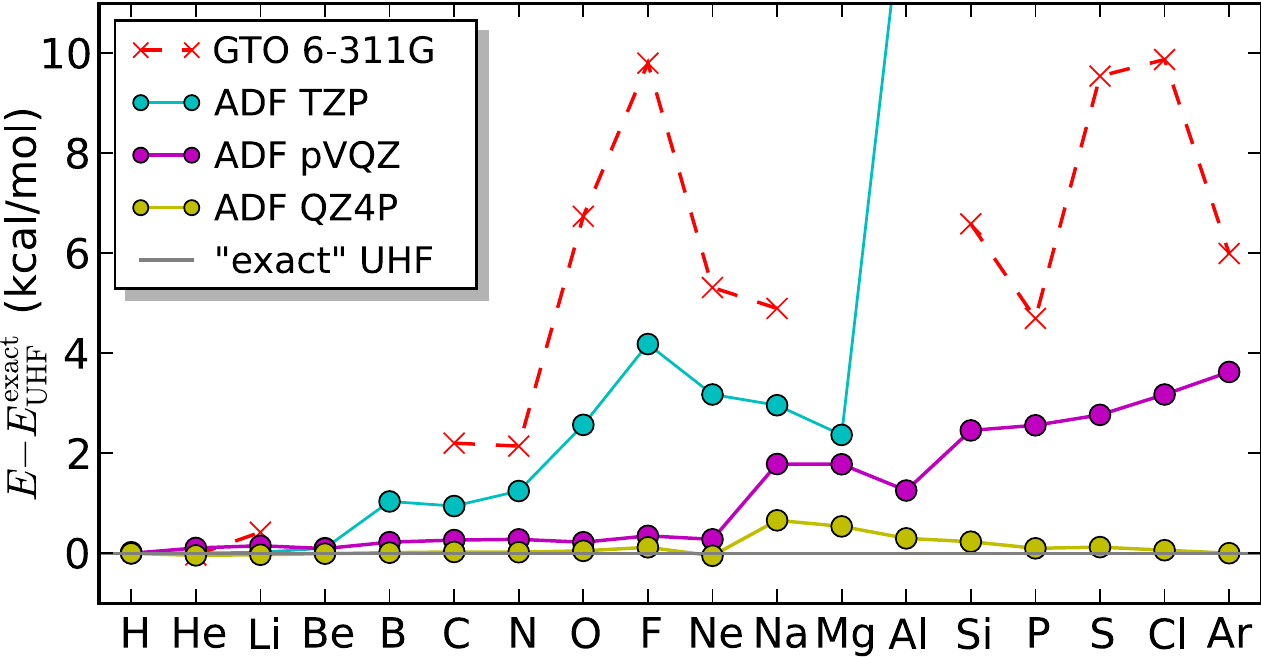}}}{\label{fig:hf-atom}Comparison
of atomic unrestricted HF (UHF) energies. The reference
energy\cite{oneill-bcefsm2005} is based on a cc-pV5Z-h GTO basis set and
assumed to be exact within the published precision of 0.1~kcal/mol.}

\tmtablesmall{\begin{tabular}{llcc}
  {\emph{type}} & {\emph{basis}} & $N_{\operatorname{bas}}$ & $T_{\operatorname{cpu}}$/step\\
  \hline
  STO & DZP & 60 & 359 $\mu s$\\
  & TZP & 74 & 391 $\mu s$\\
  & pVQZ & 160 & 531 $\mu s$\\
  & QZ4P & 212 & 643 $\mu s$\\
  &  &  & \\
  {\emph{type}} & {\emph{cusp corr.}} & $N_{\operatorname{bas}}$ & \\
  \hline
  GTO & none & 114 & 468 $\mu s$\\
  & gpcc & 114 & 486 $\mu s$\\
  & Gaussian\cite{ma-sfaectgo2005} & 114 & 512 $\mu s$
\end{tabular}}{\label{tbl:timing}Timing of VMC steps for the $\mathrm{C_2
H_6}$ molecule. The STO/TZP basis set is of comparable precision to GTO and
significantly more efficient to evaluate. For equal size of basis set, GTO
without cusp correction performs as well as STO. The two types of cusp
correction implemented in CASINO each add some computational overhead. The
general purpose cusp correction (gpcc)\cite{gpcc} is a scheme that adds a
correction function to each molecular orbital of arbitrary type, while the
Gaussian cusp correction\cite{ma-sfaectgo2005} replaces the orbital close
to nuclei by the exponential of a polynomial. For larger systems the
computational cost is expected to grow linearly with the basis set size,
promising up to a 45\% performance gain for equivalent precision.}

\section{Slater-type orbitals}

Slater-type orbitals (STO) were an important tool in quantum mechanics long
before the availability of computational tools in physics and
chemistry.\cite{slater-asc1930} Inspired by the analytic solution of the
hydrogen atom, STO basis sets were the first choice for a number of important
approximate studies in the early years of quantum chemistry. With the arrival
of computers, however, it turned out that Gaussian basis functions
(GTO)\cite{boys-ewfiagmocftssoams1950} allow a far simpler efficient
implementation due to the possibility of factorizing Gaussian functions in
Cartesian coordinates and the simplicity of evaluating multi-center integrals.
They became the standard in quantum chemistry to the point that chemists
typically discuss ``basis sets'', implicitly referring to a GTO basis.

\tmfloatsmall{\resizebox{80mm}{!}{\includegraphics{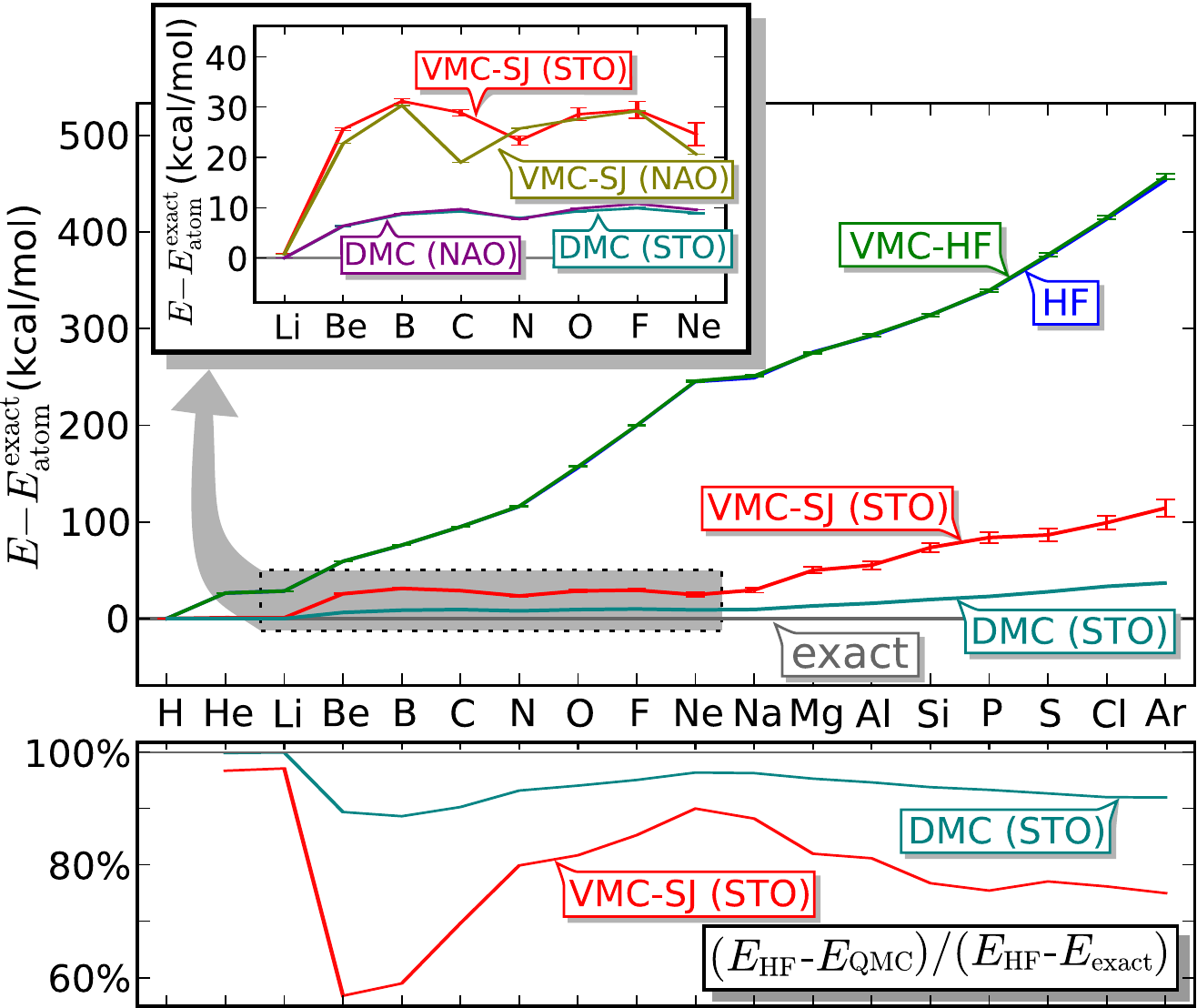}}}{\label{fig:dmc-atom}Atomic
total energies computed within HF (ADF/QZ4P) and QMC compared with the exact
results.\cite{chakravorty-gcefaiw3t1e1993} The VMC run with the HF wave
function reproduces the HF energy to within statistical error bars
(demonstrating the negligible effect of the cusp correction on the energy for
an accurate basis set). Optimized Slater-Jastrow (SJ) wave functions recover
roughly 60-85\% of the correlation energy within VMC and 90-95\% within DMC.
The results agree very well with an earlier study based on numerical atomic
orbitals (NAO) on a radial grid.\cite{brown-eotfrafqmc2007}}

While the relative merits of the GTO and STO representations of orbitals in
quantum chemistry are still under debate,\cite{shavitt-aebsptg2004}
attempts have been made to reduce the computational effort of working with STO
basis
sets,\cite{silverstone-oteotoaciwnsto1966,baerends-smhcitcp1973,allouche-nsoc1976,guseinov-comiosoxcooiwiannsoucosoef2002}
leading to the development of several STO-based electronic structure
codes.\cite{bouferguene-sasopfmesd1996,rico-npfmcwso2001,velde-cwa2001} Of
these, the only code active development and available for general use is
ADF,\cite{velde-cwa2001} offering a full state-of-the-art implementation of
Hartree-Fock (HF) and density functional theory (DFT) electronic structure
calculations for molecules and, via its sister program BAND, for periodic
systems.

\begin{figure*}
  \resizebox{165mm}{!}{\includegraphics{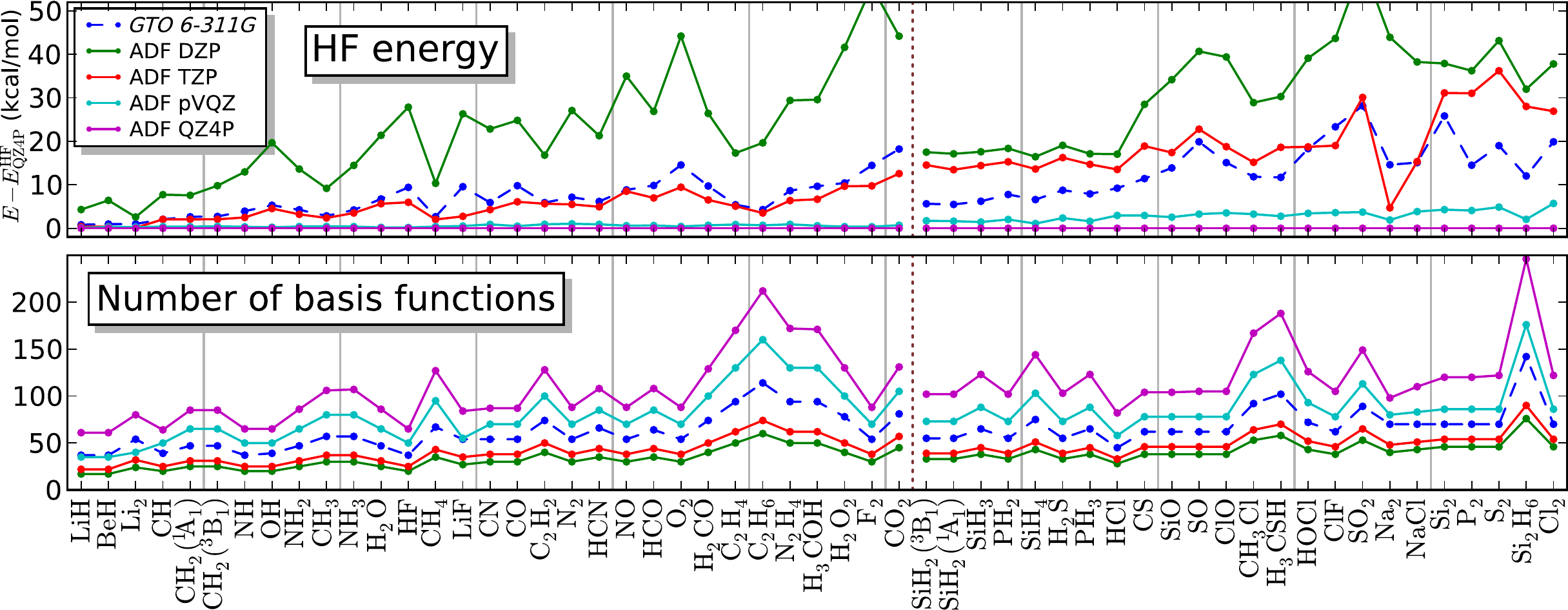}}
  \caption{\label{fig:HF-G2}Comparison of basis set errors in the HF energies
  for various STO basis sets. The QZ4P basis is used as a reference. Judging
  by Fig.~\ref{fig:hf-atom}, the remaining error is less than 1~kcal/mol for
  each second-row atom. The GTO 6-311G basis is roughly equivalent in
  precision to the STO TZP basis set which typically contains about 45\% more
  basis functions. Geometries and GTO reference data were the same as those
  used in {Ref.~\onlinecite{ma-sfaectgo2005}}~(see EPAPS material\cite{epaps}).
  The dashed line separates the molecules containing only first-row atoms from
  those also containing second-row atoms.}
\end{figure*}

Quantum Monte Carlo (QMC) methods have very different computational
requirements from conventional non-stochastic electronic structure methods.
Without the need to perform analytic integrations, the usual advantages of GTO
become irrelevant. Instead, the bulk of the computational cost lies in
evaluating the trial wave function and its derivatives at arbitrary positions
in space. A basis set that achieves the same precision with a more compact
representation (fewer basis functions) will gain a clear advantage.
Furthermore, STO wave functions allow an exact treatment of the Kato cusp
condition at nuclei\cite{kato-oteomsiqm1957} and do not suffer from
divergent local energies at large distance from a molecule. For these reasons,
STO basis sets have often been used in QMC.

In the past, the main disadvantage of STO basis sets in QMC calculations has
stemmed from the use of a trial wave function commonly generated from a
preliminary HF or DFT calculation; almost all suitable mainstream local basis
set electronic structure codes for finite systems -- particularly in the
quantum chemistry community -- use GTO. To exploit the advantages of STO for
QMC calculations, one could therefore use either a conversion
step\cite{manten-otaotfdqmcm2001} or optimize the orbitals directly within
VMC.\cite{umrigar-admcawvste1993} The advent of the ADF code has allowed
the generation of HF or DFT orbitals directly in an STO basis, which can then
be used in QMC calculations.

GTO basis sets have been available in the CASINO program for over a decade. We
have now implemented the additional capability to evaluate orbitals expanded
in a STO basis, allowing the use of trial wave functions generated by ADF in
VMC and DMC calculations.\cite{foulkes-qmcsos2001} In the following, we
present details of the cusp constraint used in the converter and demonstrate
the precision achievable with the combination of ADF and CASINO.

Each molecular orbital is expanded in the STO basis
\begin{eqnarray}
  \Psi \left( \mathbf{r} \right) & = & \sum_{i = 1}^{N_{\operatorname{bas}}} c_i
  \psi_i \left( \mathbf{r}-\mathbf{R}_i \right),  \label{eqn:molorb}
\end{eqnarray}
with $N_{\operatorname{bas}}$ basis functions of the form
\begin{eqnarray}
  \psi_i \left( \mathbf{r} \right) & = & Y_{l_i}^{m_i} \left( \vartheta,
  \varphi \right) r^{l_i + n_i} \mathrm{e}^{- \zeta_i r},  \label{eqn:sto-basfn}
\end{eqnarray}
with $\zeta_i > 0$ and the $Y_l^m$ are the Laplace spherical harmonics.
Abandoning orthogonality in favour of simplicity, the Laguerre polynomials
present in the analytic solutions for the hydrogen atom are replaced by $r^n$.

The centers $R_i$ of the basis functions usually coincide with the positions
$R_I$ of the $N_{\operatorname{nuc}}$ point-like nuclei $I$ of charge $Z_I$. In
principle, a single value of $\zeta$ would allow the construction of a
complete basis set by including sufficiently high orders $n$. In practice,
however, using a small number of different $\zeta$ values is a more efficient
means of improving the precision of the basis set. A further improvement in
precision can be made by including high-angular-momentum basis functions to
improve the description of polarization. In this work, we used four
general-purpose basis sets from the ADF package, in increasing size and
precision: single-$\zeta$ (SZ), double-$\zeta$ (DZ), triple-$\zeta$-polarized
(TZP) and quadruple-$\zeta$-fourfold-polarized (QZ4P).

\begin{figure*}
  \resizebox{165mm}{!}{\includegraphics{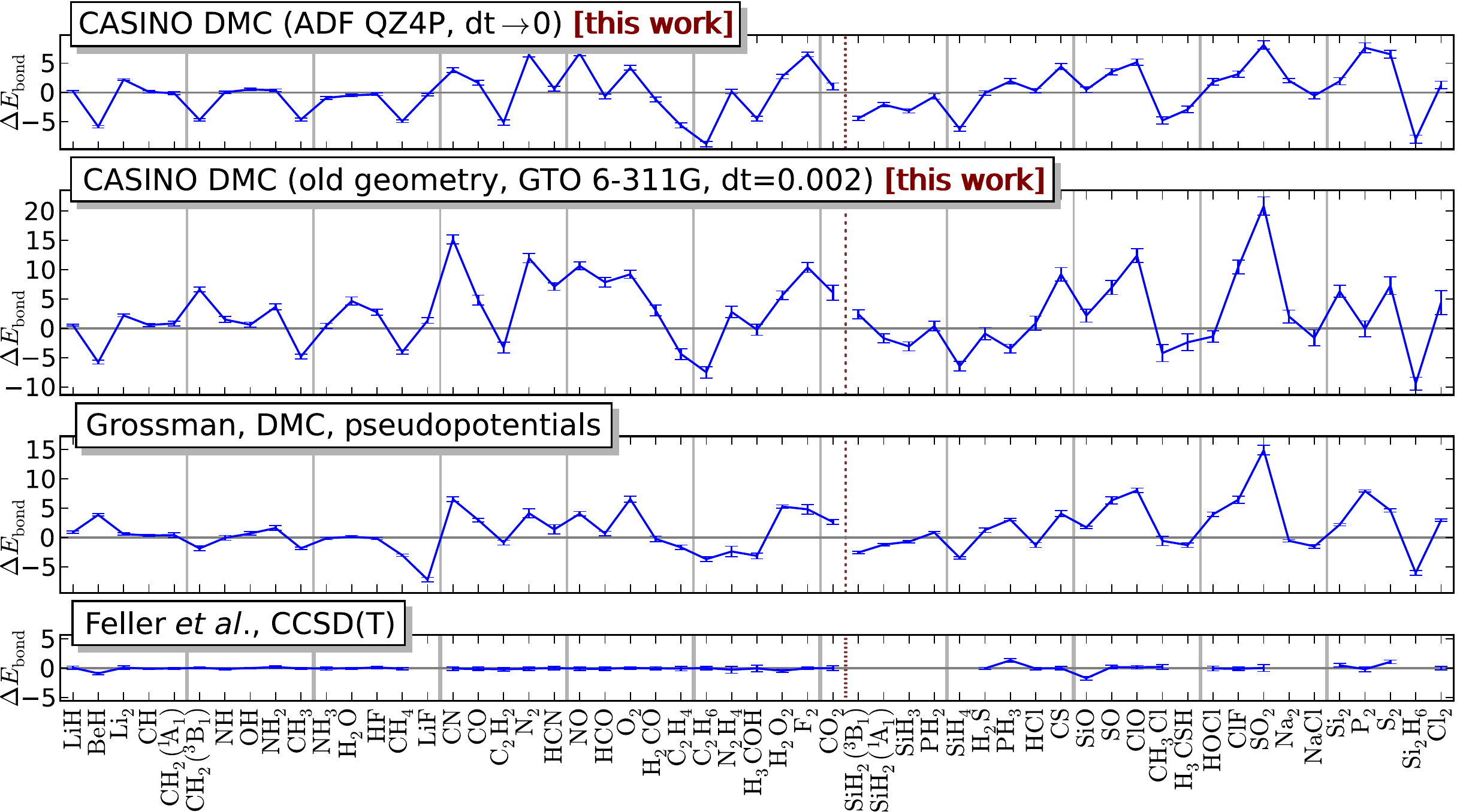}}
  \caption{\label{fig:G2-benchmark}Comparison of various calculated bond
  energies from \textit{ab initio} computations. The deviation $\Delta
  E_{\operatorname{bond}} = E_{\operatorname{bond}} - E_{\operatorname{ref}}^{\operatorname{bond}}$ from the
  experimental reference energy given in {Ref.~\onlinecite{feller-roaeftgsom1999}}
  (including zero-point motion, relativistic and spin-orbit corrections) is
  shown. Our GTO-based DMC calculations were performed with deliberately less
  effort than the STO-based ones. Grossman\cite{grossman-bqmcc2002} chose a
  DMC approach similar to ours but using pseudopotentials whereas we have used
  an all-electron approach. The excellent values obtained by Feller
  \textit{et al.}\cite{feller-asofctatpoaeams2008} are based on
  fixed-core coupled cluster computations with a careful choice of basis set
  for each molecule including core-valence corrections.}
\end{figure*}

\section{Cusp constraint}

The exact wave function of particles interacting via a Coulomb interaction
fulfills the Kato cusp condition whenever two point-like particles
coalesce.\cite{kato-oteomsiqm1957} For pairs of electrons, this condition
gives rise to dynamic correlations that can be very efficiently represented by
a Jastrow
factor.\cite{jastrow-mpwsf1955,boys-acfteawfsonwfeca1969,schmidt-cmcwfftahtn1990,drummond-jcffamas2004}
In all-electron calculations, each single-electron orbital $\Psi$ should
fulfill the cusp condition
\begin{eqnarray}
  \left. \left\langle \frac{\mathrm{d}}{\mathrm{d} r} \Psi \left( \mathbf{r} \right)
  \right\rangle_{\Omega} \right|_{\mathbf{r}=\mathbf{R}_I} & = & - Z_I
  \Psi \left( \mathbf{R}_I \right)  \label{eqn:cusp-condition}
\end{eqnarray}
in the vicinity of each point-like nucleus $I$ of charge $Z_I$ at position
$\mathbf{R}_I$, where $\left\langle \cdot \right\rangle_{\Omega}$ denotes
the spherical average around the nucleus.

In methods such as DFT or HF, having the exact cusp at the nucleus is less
important than the overall quality of the wave function, so smooth Gaussian
functions can be used to represent the wave function. QMC, on the other hand,
is based on evaluating the local energy which diverges at coalescences if the
cusp condition is not exactly fulfilled.\cite{ma-sfaectgo2005} When using
GTO-based trial wave functions in QMC, the cusp condition is typically
enforced artificially, either by modifying the GTO basis
functions\cite{kussmann-aectgbffmqmcc2007} or by directly constructing a
correction to the single-electron
orbitals.\cite{ma-sfaectgo2005,per-eccafiqmc2008}

In contrast to smooth GTO basis functions, STO-based orbitals
[Eq.~(\ref{eqn:sto-basfn})] are able to fulfill the cusp condition
[Eq.~(\ref{eqn:cusp-condition})] exactly, leading to one linear constraint per
nucleus $I$ on the coefficients $c_i$ of any molecular orbital $\Psi$. These
$N_{\operatorname{nuc}}$ constraints can be expressed as a single matrix equation
\begin{eqnarray}
  \sum_i \chi_i^I c_i & = & 0,  \label{eqn:cusp-constraint}
\end{eqnarray}
where the $N_{\operatorname{nuc}} \times N_{\operatorname{bas}}$ elements of the constraint
matrix are given by
\begin{eqnarray*}
  \chi_i^I & = & \delta_{R_i, R_I} \delta_{l_i, 0} [\delta_{n_i, 0} \left(
  \zeta_i - Z_I \right) - \delta_{n_i, 1}] -\\
  &  & \hspace*{\fill} - Z_I \left( 1 - \delta_{R_i, R_I} \right) \psi_i
  \left( \mathbf{R}_i -\mathbf{R}_I \right) .
\end{eqnarray*}
In principle, these linear constraints can be enforced during a HF
computation,\cite{galek-howotncc2005} but this is rarely implemented in
electronic structure codes. Instead a wave function originating from a code
such as ADF can be cusp-corrected by a linear projection. To restrict the
effect of the cusp correction to the vicinity of the nuclei, we fix all
coefficients except those of the narrowest 1s basis function on each nucleus,
and adjust the coefficients of the $N_{\operatorname{nuc}}$ remaining orbitals to
fulfill Eq.~(\ref{eqn:cusp-constraint}). As the cusp conditions on different
nuclei are nearly independent, this linear problem is always well-conditioned
and has a unique solution.

Fig.~\ref{fig:cusp-correction} demonstrates the effect of this cusp correction
scheme on the orbitals and local energies. The single-$\zeta$ basis set
clearly does not leave sufficient freedom for the correction and the single 1s
basis function is severely distorted. For larger basis sets, however, the
singularity in the local energy is cleanly removed with negligible distortion
of the orbitals away from the cusp.

\begin{figure*}
  \resizebox{165mm}{!}{\includegraphics{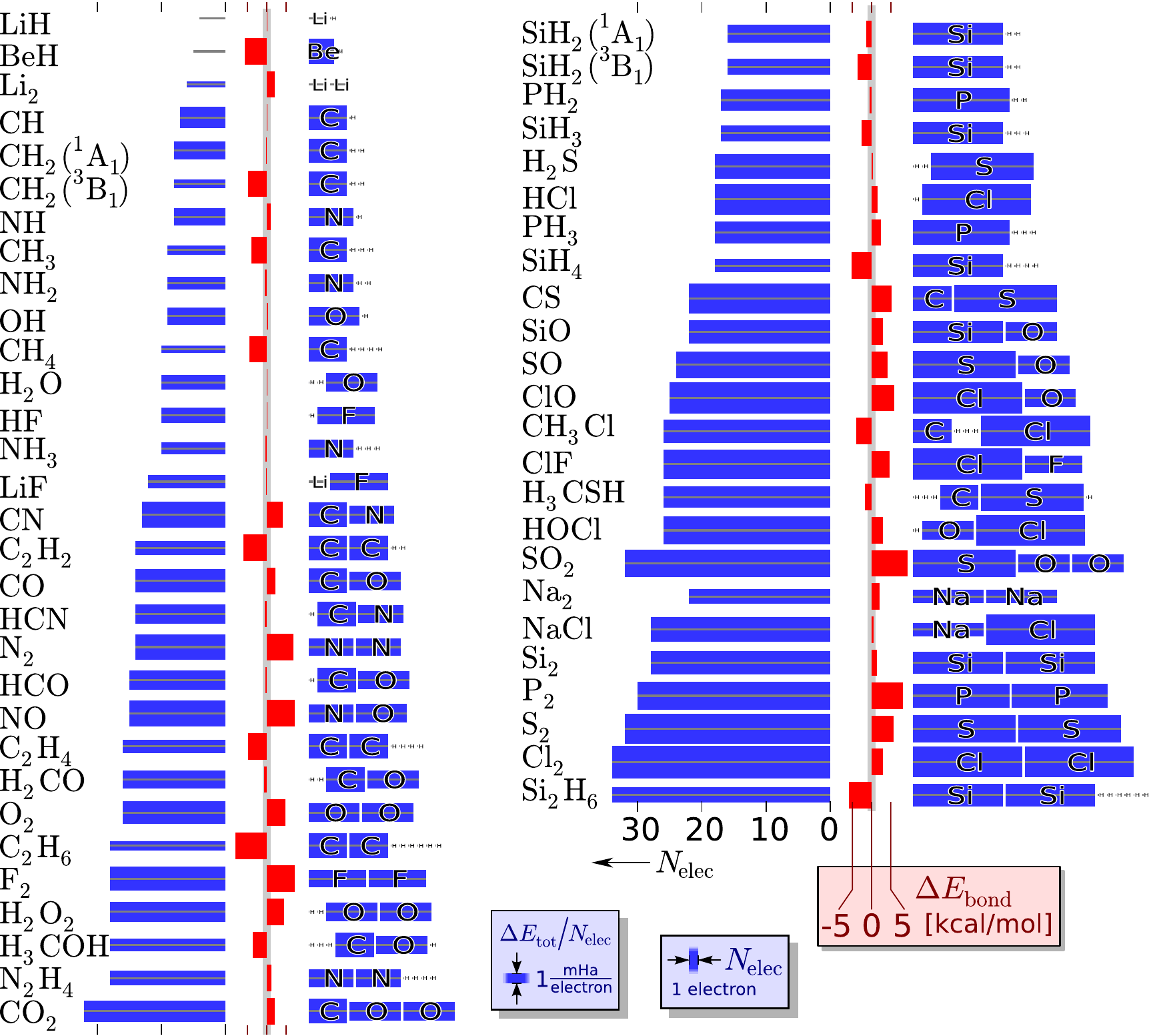}}
  \caption{\label{fig:error-cancellation}Visualization of the error
  cancellation in bond energies. The horizontal bars on the left and right of
  each reaction correspond to the error in the total energy of each species
  $\Delta E_{\operatorname{tot}} = E_{\operatorname{tot}}^{\operatorname{dmc}} -
  E_{\operatorname{tot}}^{\operatorname{ref}}$, where the reference total energies are based
  on experimental atomization energies, atomic total energies and theoretical
  corrections.\cite{oneill-bcefsm2005} Within the statistical precision,
  these errors are due almost entirely to the fixed node approximation. The
  vertical extent of each bar is the error per electron, which is in the range
  of 1--3~mHa/electron except for {\molecule{H}}, {\molecule{H_2}},
  {\molecule{He}}, {\molecule{Li}}, {\molecule{LiH}}, {\molecule{BeH}}, and
  {\molecule{Li_2}} for which the nodal surface is exact or nearly so. The
  horizontal extent is the number of electrons. The difference in the areas on
  the right and left side of each reaction represents the error in the bond
  energy, which is shown in the center.}
\end{figure*}

\section{Atomic total energies}

We computed the HF energies of the first- and second-row atoms with ADF (see
Fig.~\ref{fig:hf-atom}) to compare the quality of available basis sets. We
found the ADF/QZ4P basis-set error to be below 0.1~kcal/mol for the first row
atoms and below 1~kcal/mol for the second row atoms. The corresponding DMC
energies are expected to be less sensitive to errors in the orbital basis than
HF energies. For comparison, GTO-based HF energies were computed with the
CRYSTAL program,\cite{crystal2003} using a 6-311G basis set, also displayed
in Fig.~\ref{fig:hf-atom}.

To evaluate the combined approach of using ADF and CASINO, the total energies
of the first- and second-row atoms including correlation effects are compared
with exact reference values in Fig.~\ref{fig:dmc-atom} as well as with earlier
QMC data using wave functions defined on a radial
grid.\cite{brown-eotfrafqmc2007} A VMC calculation using the
Slater-determinant wave function reproduced the HF energy, confirming that
enforcing the cusp constraint preserves the overall quality of the wave
function.

For each atom, we optimized a Jastrow factor\cite{jastrow-mpwsf1955}
consisting of electron-electron, electron-nucleus and
electron-electron-nucleus terms. The Jastrow factor was expressed as a power
expansion in the inter-particle distances with the parameters: $C = 3$, $N_u =
N_{\chi} = 12$, $N_f^{\operatorname{en}} = 2$, $N_f^{\operatorname{ee}} = 3$ (see Eqs. 19, 20,
and 21 of Drummond \textit{et al.}\cite{drummond-jcffamas2004}). For the
VMC optimization, we minimized the mean absolute deviation of the energy from
the mean energy over a set of $50000 / N_{\operatorname{elec}}$ configurations,
performing five consecutive optimization iterations. This conservative choice
of parameters allowed a reliable, automated optimization procedure. The
resulting Slater-Jastrow (SJ) wave functions recover 60--85\% of the
correlation energy.

DMC calculations using the optimized SJ wave functions recovered 90--95\% of
the correlation energy for Be and heavier atoms. (Time step
errors\cite{umrigar-admcawvste1993} were eliminated by linear
extrapolation.) H and He do not have a nodal surface, so DMC produces the
exact energy. The HF nodal surface for Li is extremely good, and DMC recovers
more than 99.5\% of the correlation energy in this case. The total energies of
the first row atoms show excellent agreement with previous QMC
results.\cite{brown-eotfrafqmc2007} This indicates that the remaining error
is due to the fixed-node approximation and presents the limit of the single
determinant SJ method which can only be bettered by improving the nodal
surface, e.g. by using backflow\cite{ros-ibtiqmcc2006} or multi-determinant
wave functions.\cite{buenda-qmcgseftalta2009}

\tmfloatsmall{\resizebox{80mm}{!}{\includegraphics{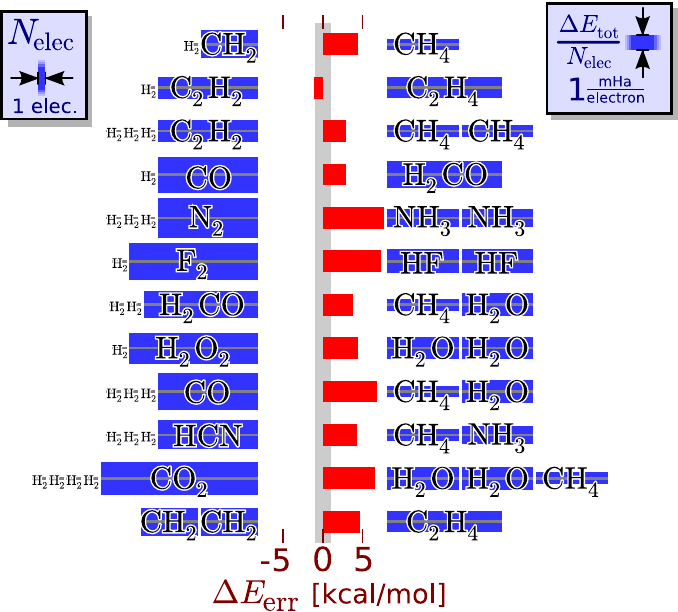}}}{\label{fig:error-cancellation-helgaker}The
same visualization scheme as introduced in Fig.~\ref{fig:error-cancellation},
applied to some of the chemical reactions studied in
{Ref.~\onlinecite{klopper-hacomes1999}}. In most cases, the nodal surface of the
unsaturated molecule on the left is described significantly worse by single
determinant wave function than the saturated molecules on the right, leading
to a systematic overestimation of reaction energies as reported previously
.\cite{manten-otaotfdqmcm2001,korth2009}}

\section{The G2 set of molecules}

Further benchmarking was performed for the bond energies of the 55 molecules
of the G2 set.\cite{curtiss-gtfmeofasc1991} The bond energy
$E_{\operatorname{bond}}$ is the difference between the molecular and atomized total
energies, not including the zero point motion of the
nuclei\cite{oneill-bcefsm2005} or any further
corrections.\cite{feller-asofctatpoaeams2008}

First of all we used the G2 set to analyze the quality of the various basis
sets by comparing the molecular HF energies (see Fig.~\ref{fig:HF-G2}). Based
on this data and the atomic data, we chose the QZ4P basis set for all further
work as it systematically gives the smallest basis set error.

To enable a direct comparison of results, the HF calculations
(Fig.~\ref{fig:HF-G2}) were performed using the same molecular geometries as
the previous GTO calculations reported in {Ref.~\onlinecite{ma-sfaectgo2005}}. For
all other computations we used the more precise geometries from
{Ref.~\onlinecite{feller-asofctatpoaeams2008}} where available, and for the
remaining molecules we used those from {Ref.~\onlinecite{oneill-bcefsm2005}} which
were provided by the authors.\cite{epaps} The geometry of
{\molecule{SiH_2}} in the triplet state was obtained from
{Ref.~\onlinecite{kalemos-sacs2004}}.

For each molecule, we optimized a Jastrow factor with the same set of terms
and parameters as used for the atoms. On average, we recovered 82\% of the
correlation energy within VMC.

The SJ wave functions thus obtained were then used as the trial wave functions
for DMC. For each system, two runs at timesteps $\operatorname{dt} = 0.01$ and
$\operatorname{dt} = 0.001$ were performed, allowing a linear extrapolation $\operatorname{dt}
\rightarrow 0$. We used a total of 40 CPU hours per electron (Intel XEON, 3
GHz) for each molecule, reaching a statistical precision of about 25~$\mu
\operatorname{hartree}$ per electron.

For comparison, the bond energies computed using different methods are shown
in Fig.~\ref{fig:G2-benchmark}. The mean absolute deviation (MAD) of our
STO-based bond energies from the experimental reference values is
3.2~kcal/mol. This is slightly larger than the deviation of 2.9~kcal/mol found
in the pseudopotential-based DMC study of
Grossman,\cite{grossman-bqmcc2002} however, those values excluded the
relativistic and spin-orbit corrections which shift individual reference
values by up to 2 kcal/mol and would increase the MAD to 3.1~kcal/mol. Indeed,
the MAD between Grossman's pseudopotential results and our all-electron
results is just 2.0 kcal/mol, showing the strong correlation between the
errors in the two sets of results. Our GTO-based all-electron DMC
calculations, which were deliberately performed without time step
extrapolation or any kind of fine tuning of the basis sets or other
computational parameters, gave a MAD from the reference values of
5.1~kcal/mol. This larger MAD shows that time-step extrapolations and a
careful choice of basis sets is important in obtaining accurate
single-determinant SJ DMC results.

Overall the obvious correlations between the errors of the three independent
DMC-based attempts clearly suggest that a further systematic improvement can
only be achieved by going beyond the fixed-node DMC approach with
single-determinant SJ trial wave functions. This confirms the finding of
Grossman that the fixed-node approximation dominates the remaining error.

\section{Error cancellation}

The variational principle guarantees that the fixed node approximation leads
to an overestimate of the total energy for each molecule and atom. The bond
energy, being a difference of total energies, therefore shows significant
error cancellation. Studying the bond energies in Fig.~\ref{fig:G2-benchmark}
reveals only very limited systematics about the sign and the magnitude of the
errors in the bond energies. The picture becomes much clearer when we directly
compare the errors in the total energies for the molecules and their
constituent atoms. We found that the nodal surface error lies in the range of
1--3~mHa/electron for each atom and molecule in our test set, except for a
very limited set of species for which the nodal surface is exact or nearly so.
This allows us to visualize the error cancellation in a very compact and
intuitive manner (Fig.~\ref{fig:error-cancellation}).

We observe a limited number of cases where both molecule and constituent atoms
are described well, leading to an accurate bond energy ({\molecule{LiH}},
{\molecule{Li_2}}). In some cases, all species involved show similar nodal
errors per electron, leading to strong error cancellation (e.g.
{\molecule{CO}}, {\molecule{CO_2}}, {\molecule{Na_2}}, {\molecule{Si_2}}). In
many other cases, however, the quality of the wave function is very different
for the various species, which may lead to a large net error (e.g.
{\molecule{NO}}, {\molecule{SO_2}}), although the errors may also largely
cancel (e.g. {\molecule{LiF}}, {\molecule{NaCl}}).

For most molecules containing hydrogen and carbon (e.g. {\molecule{CH_3}},
{\molecule{CH_4}}), one can observe that the wave function of the molecule is
described significantly better than that of the carbon atoms, leading to a
systematic over-estimation of the bond energy. (Even though the absence of a
nodal surface error for the hydrogen atom might intuitively suggest the
opposite.) A similar effect can also be seen in the visualization of chemical
reaction energies (Fig.~\ref{fig:error-cancellation-helgaker}). Here, one can
observe that fully hydrogenated molecules are typically described better by
the fixed node approximation than molecules containing double or triple bonds.

Finally, one can observe that second row atoms and their molecules typically
show significantly larger nodal surface errors than first row atoms. The
errors in bond energy, however, are not necessarily larger, indicating that
these nodal surface errors arise mainly from the core electrons and that they
cancel in energy differences.

\section{Conclusions}

To conclude, we have demonstrated the accuracy of STO trial wave functions
generated by the ADF software packages in VMC and DMC calculations using the
CASINO program. Using the QZ4P basis set from ADF, the basis set errors are
below 0.1~kcal/mol for first-row atoms and below 1~kcal/mol for second-row
atoms. DMC calculations for the G2 set of molecules recovered on average 95\%
of the correlation energy. Due to partial error cancellation, the atomization
energies could be reproduced to a mean absolute deviation from the
experimental values of 3.2~kcal/mol.

The errors in the total energies of individual molecules and atoms, which
originate -- apart from statistical errors -- almost entirely from the fixed
node approximation, were then used to analyse and visualize the error
cancellation in atomization and chemical reaction energies. While we find that
the nodal error from the core electrons in the second row atoms largely
cancels out, other error cancellations seem more coincidental than systematic.

As our bond energies are of similar quality to those obtained previously in
pseudopotential calculations, we may assume that we have reached the limit in
accuracy that is possible with a nodal surface obtained from single Slater
determinant wave functions. Systematic studies of the nodal surface of
multideterminant wave functions\cite{bressanini-aionsatcotwf2005} indicate
that significant improvements can be achieved with reasonable effort. Using
backflow functions\cite{ros-ibtiqmcc2006} or geminal wave
functions\cite{casula-gwfwjcafata2003} should also lead to higher accuracy
QMC results while retaining its excellent performance and scaling behavior.

The molecular geometries and the full set of results are provided in
electronic form, available from EPAPS.\cite{epaps}

We acknowledge the authors of the ADF software for permission to use the code,
together with excellent support by S.~van~Gisbergen and E.~van~Lenthe. We
thank D.~O'Neill and P.~Gill for providing molecular coordinates and M. Korth
for valuable discussions. This work was funded by the DAAD and the Engineering
and Physical Sciences Research Council (EPSRC) of the United Kingdom. M.D.T.
acknowledges financial support from the Royal Society. The computations were
performed using the facilities of the University of Cambridge High Performance
Computing Service.

\end{document}